\documentstyle[aps,preprint,epsf]{revtex}

\begin{document}

\title{Liquid-Gas Phase Transition in Nuclear Equation of State}
\author{S.J. Lee}
\address{Department of Physics and Institute of Natural Sciences \\
   Kyung Hee University, Yongin, Kyungkido,449-701 KOREA}
\author{A.Z. Mekjian}
\address{Department of Physics and Astronomy, Rutgers University \\
   Piscataway, New Jersey 08855-0849  }

\maketitle

\begin{abstract}
A canonical ensemble model is used to describe a caloric curve
of nuclear liquid-gas phase transition.
Allowing a discontinuity in the freeze out density from one spinodal
density to another for a given initial temperature,
the nuclear liquid-gas phase transition can be described as first order.
Averaging over various freeze out densities of all the possible initial
temperatures for a given total reaction energy, the first order
characteristics of liquid-gas phase transition is smeared out to
a smooth transition.
Two experiments, one at low beam energy and one at high beam energy
show different caloric behaviors and are discussed.
\end{abstract}



Recent experiments of heavy ion collisions at low and high beam energy
show caloric curves for nuclear liquid-gas phase transition that
are different \cite{prldata,predata}.
Even if there is some uncertainty concerning the extraction of temperatures
in heavy ion collision \cite{campi}, low energy collision shows
some indication of a sharp first order phase transition \cite{prldata}
while higher energy collision shows a much smoother phase
transition \cite{predata}.
The data of higher energy collision is also used in extracting the
critical exponents of nuclear system \cite{crit}.

The investigations of nuclear phase transition through heavy ion
collision are based on two basic assumptions which are not confirmed:
One is that we can apply equilibrium thermodynamics
for such a small system of few hundreds constituents at the most.
Another is that a thermalized uniform system is formed in heavy ion
collision before the multigfragmentation takes place.
Based on these assumptions,
a canonical ensemble model describing nuclear multifragmentation phenomena
have been developed \cite{prcano}.
Another assumption of this model is that the fragment distribution
is directly related with the state of thermalized uniform system before
its break up.
In this paper we will use this model to describe the caloric curve
of nuclear liquid-gas phase transition.
This model may also determine a beam energy required for the study of
critical behavior.

We first summarize some results of the canonical ensemble model
of Ref.\cite{prcano}
which we will use in this paper. Further details and derivations can
be found in that reference and references therein.
In Ref.\cite{prcano}, when each cluster of size $k$ is weighted
by $x_k = x X_1^k$, the canonical ensemble averaged number of clusters
of size $k$ is given by
\begin{eqnarray}
 <n_k> = \frac{x}{k} \frac{A!}{(A-k)!} \frac{\Gamma(x + A - k)}{\Gamma(x + A)}
                   \label{nk}
\end{eqnarray}
for a system of $A$ nucleons.
The result is independent of $X_1$, i.e., the weighting factor assigned
to each nucleon. The mean cluster distribution thus depends only
on $x$ which is the same for each cluster.
As $x \to 0$, the system forms one large cluster.
As $x \to \infty$, only nucleons are present.

The $x_k$ can be obtained \cite{prcano} by weighting each cluster
with its free energy as
$x_k = A x_\rho e^{-T \int^T d T E_k(T)/T^2}$ with
\begin{eqnarray}
 x_\rho = \frac{g}{h^3}(2\pi M_B)^{3/2} \frac{1}{\rho_0}
        \left[ \frac{\rho_0}{\rho} - 1 \right]
   .     \label{xv}
\end{eqnarray}
Here $M_B$ is the nucleon mass, $g = 4$ is the degeneracy factor.
The freeze out density $\rho$ is the density when
there are no further break up or clusterization occuring.
An equilibrium temperature $T$ determines the energy shared between
the liquid phase part (internal energy of clusters)
and the gas phase part (thermal motion of clusters).
If we approximate
the energy for a cluster of size $k$ at temperature $T$ as
\begin{eqnarray}
 E_k(T) = \frac{3}{2} T
    - \left[ a_B - \frac{1}{\epsilon_o} \left(\frac{T_0 T}{T + T_0}\right)^2
             \right] k                  ,    \label{ekt}
\end{eqnarray}
then the common weight factor of each cluster becomes $x = A x_\rho T^{3/2}$.
Here $a_B$ is the average binding energy at
zero $T$, $\epsilon_0$ is the level spacing parameter and
the $T_0$ is a cut-off temperature for internal excitation of the cluster.
The $a_B \approx 16$ MeV for a nuclear matter and
$a_B \approx \epsilon_0 \approx 8$ MeV for finite nuclei.

Substracting the thermal energy $\frac{3}{2} T$ of the center of mass,
the canonical ensemble averaged total energy per nucleon
is given by 
\begin{eqnarray}
 E(T) &=& \left[ \sum_{k=1}^A E_k(T) <n_k> - \frac{3}{2} T \right]/A
      .   \label{et}
\end{eqnarray}
The excitation energy per particle $E^*(T) = E(T) + a_B$ is
\begin{eqnarray}
  E^*(T) &=& \frac{3}{2} T \sum_{r=1}^{A-1}
         \frac{1}{A + r x_\rho^{-1} T^{-3/2} }
     + \frac{1}{\epsilon_0} \left(\frac{T_0 T}{T+T_0}\right)^2
              .      \label{etpa}
\end{eqnarray}         
The low $T$ limit of Eq.(\ref{etpa}), $E(T \to 0) \approx T^2/\epsilon_0$,
gives the Fermi liquid limit 
and the high $T$ limit of Eq.(\ref{etpa}),
$E(T \to \infty) \approx \frac{3}{2} T$,
gives the ideal gas limit \cite{prldata}.
These show that the caloric curve is independent of the
freeze out density for low and high temperatures.

Eq.(\ref{etpa}) shows that the phase transition diagram ($E$--$T$
caloric curve) of a nuclear system
is determined once the freeze out density $\rho$ is known.
Eq.(\ref{etpa}) is a smooth function of temperature $T$ for fixed
$\rho$ with a constant $x_\rho$. If the nuclear density $\rho$
changes smoothly as the $T$ changes, then $E(T)$ will have
an extra $T$ dependence than the explicit dependence of Eq.(\ref{etpa})
but it still changes continuously.
However, if the density $\rho$ changes suddenly at some temperature $T$,
then the excitation energy Eq.(\ref{etpa}) has a discontinuous jump at the
temperature through the discontinuity of $x_\rho$.
The gap of this jump is the latent heat of the liquid-gas phase transition.
In a spinodal instability region, a system can not keep its uniformity
and thus breaks up into a mixture of gas and liquid.
The densities at the liquid and gas spinodal points
of an isothermal curve for a given $T$ gives a jump in the
freeze out density from liquid to gas phase.

\begin{figure}[tb] 
 \centerline{ \epsfxsize=3.0in \epsfbox{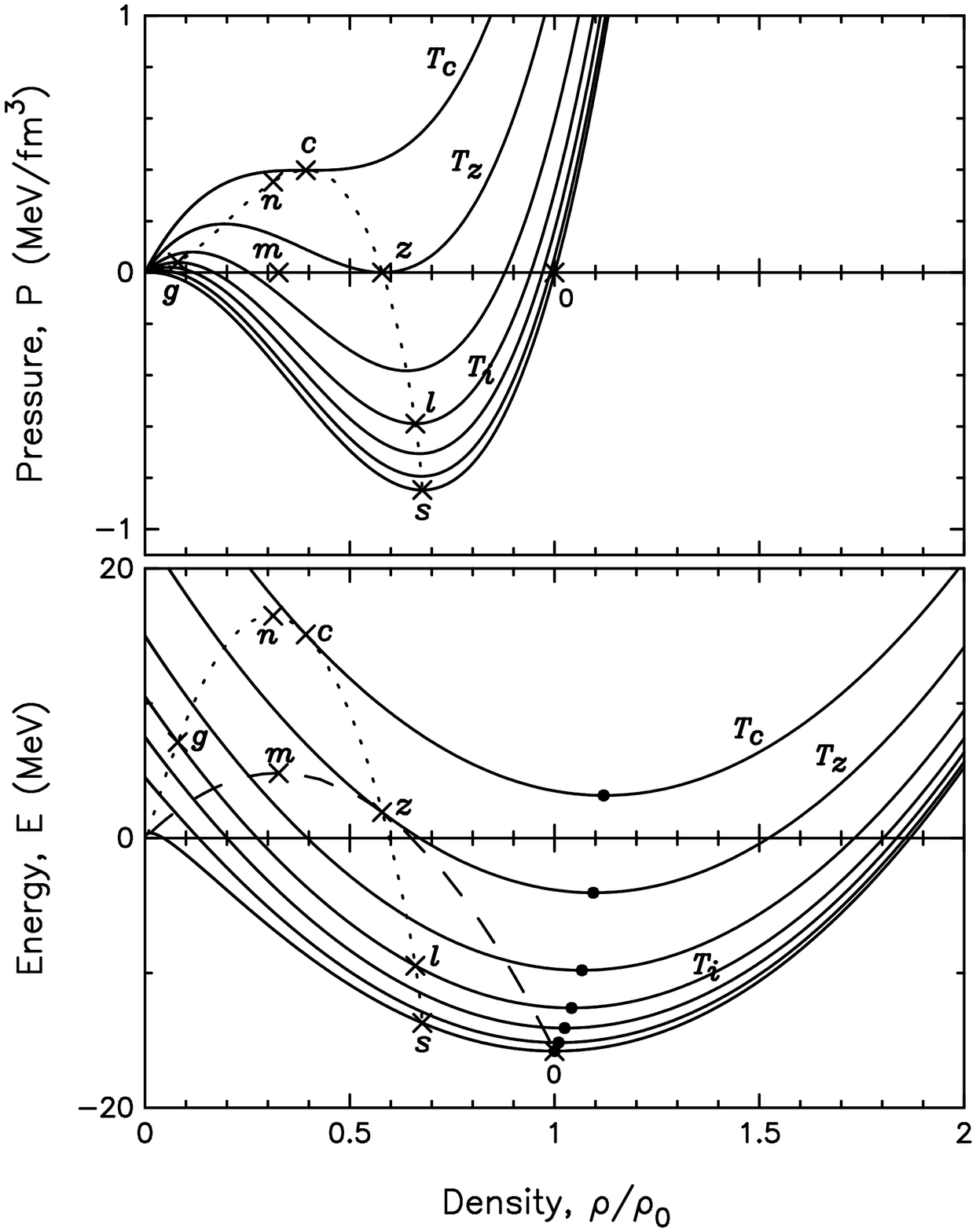}  }
 \vspace{0.5in}
\caption{  \label{eosfig}
Equation of state of nuclear matter  \protect\cite{prcano} for temperature
$T$ of, from bottom to top, 0, 3, 5, 7, 10, 15.16 ($T_z$ at which the minimum
pressure is zero), and 20.95 (which is the critical temperature $T_c$) in MeV.
Also shown are the spinodal points (dotted line), 
the points with zero pressure (dashed line), 
and the points of minimum energy (filled circles) 
for each temperature $T$.
See the text for variously marked points of cross.}
\end{figure}

For a thermally equilibrated uniform system,
the density is determined once we know the temperature and pressure or
the temperature and energy as shown in Fig.\ref{eosfig}.
If we add energy to the system keeping the temperature fixed,
then the density and the pressure will be changed to new values
according to the isothermal curve of Fig.\ref{eosfig}.
On the other hand, if we add energy keeping the pressure (density) fixed,
then the temperature and the density (pressure) will be changed.
To maintain non-zero pressure for a finite system, we need some mechanism
which can controll the volume such as infinite potential wall.
However, since such a mechanism does not exists, the system should move
toward the $P = 0$ point accompanied by a corresponding temperature and
density change.
Thus if we add energy adiabatically to a finite nucleus, then the system
changes along an isobaric path with zero pressure.
Such a system with conserved total energy $E$ will finally settled down
with some temperature $T$ and zero pressure $P$. This process is similar
to the equilibration of (adiabatically) heated water at atmospheric pressure.

As shown in Fig.\ref{eosfig} 
the pressure is always positive for $T > T_z$ where $T_z$ is
the temperature at which the minimum pressure is zero.
In the lower figure of Fig.\ref{eosfig}, the $P = 0$ isobaric curve
is shown by a dashed line.
Nuclear matter is always in spinodal stability region for $T > T_c$
where $T_c$ is the critical temperature.
The spinodal curve is shown by dotted line in this figure.
Also shown in Fig.\ref{eosfig} are the critical point $c$,
the point $n$ of maximum energy $E_n$ of spinodal curve,
the point $m$ of maximum energy $E_m$ of zero pressure isobaric curve,
and the point $z$ with energy $E_z$ which is a spinodal point with $P = 0$.

\begin{figure}[tb] 
 \centerline{ \epsfxsize=3.0in \epsfbox{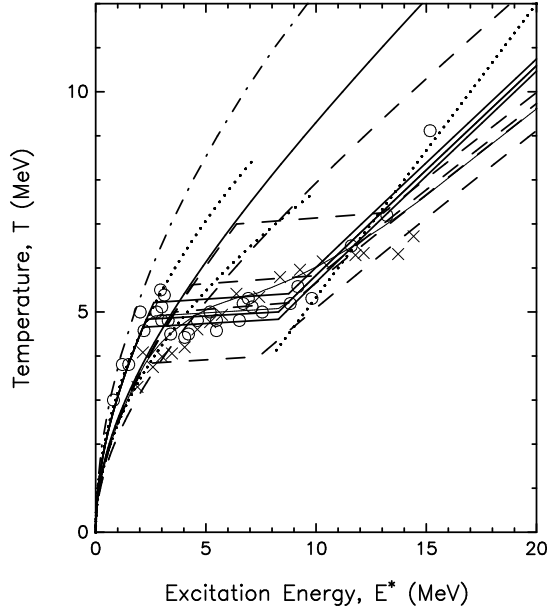}  }
\caption{  \label{phstrnf}
Phase transition caloric curve. The open circles are the data of
Ref.\protect\cite{prldata}
and the crosses are the data of Ref.\protect\cite{predata}.
The solid curves are the fit of Eq.(\protect\ref{etpa}) with spinodal
densities at $T$ of, from bottom to top, 0, 1, 2 MeV, and $T_c$
with $a_B = \epsilon_0 = 10$ MeV and $T_0 = T_c$ for $A = 50$.
The dashed curves are the fit of Eq.(\protect\ref{etpa}) with spinodal
densities at $T$ of, from bottom to top, 0, 4, 5, 7 MeV, and $T_c$
with $a_B = \epsilon_0 = 6$ MeV and $T_0 = T_c$ for $A = 110$.
The thin solid curves are the fits as discussed in the text.
The dotted curves are, from the top to bottom,
$T = \protect\sqrt{10 E^*}$, $T = \protect\sqrt{6 E^*}$,
and $T = (2/3)(E^* - 2.0)$ in MeV,
as in Ref.\protect\cite{prldata}.
The dash-dotted line is the zero pressure curve of Fig.\protect\ref{eosfig}
with $a_B = 15.77$ MeV.
  }
\end{figure}

When nuclear matter is heated adiabatically, the caloric curve should
follow the zero pressure isobaric curve shown by dash-dotted line
in Fig.\ref{phstrnf} until it reaches the point of $T = T_z$
where the zero pressure state gets into
the spinodal instability region, i.e., while $E < E_z$.
For Fig.\ref{eosfig}, 
where $a_B = 15.77$ MeV, $T_z = 15.16$ MeV and $E^*_z = 17.7$ MeV.
If the energy $E$ becomes higher than $E_z$,
the system can not be in a liquid phase as a whole with $P = 0$.
The only possibilities 
are then a uniform gas phase of constituent particles
or a mixture of clusters of liquid phase and constituent particles.
If a thermalized system belongs to the spinodal instability region, 
then the system would fragment into clusters.
For dynamical equilibrium (stability) of a cluster, each cluster
should have zero internal pressure.
For thermal equilibrium, the $T$ associated with thermal
motion of any cluster and its internal excitation should be the same.
For a thermally and dynamically equilibrated system,
the mean fragment distribution and the ensemble averaged total energy
are given by Eqs.(\ref{nk}) and (\ref{etpa}) respectivley.
For $E > E_n$, 
the system should expand to infinite volume approaching an ideal gas phase
of nucleons since it stays outside the spinodal instability region with
positive pressure.

A system produced in heavy-ion collision can break up into pieces
before it reaches thermal equilibrium.
During such a collision, the nucleons and the clusters
broken off take some energy as separation energy and thermal energy
and the remnant equilibrates thermally at lower $T$ and $E$.
For a heavy ion collision with beam energy per nucleon of $E_{beam}$,
the thermalized remnant would have $E = E^* - a_B$ and $T$ within the
limits $0 \le E^* \le E_{beam}$ and $0 \le T \le T_{max}$,
where $T_{max}$ is the maximum $T$ which can be reached
by a uniform system with energy $E$,
i.e., the minimum energy of the $T_{max}$ isothermal curve is $E$.

Suppose we have a thermalized uniform system with energy $E$ and
temperature $T_i$ satisfying EOS of Fig.\ref{eosfig}.
For a given value of $E$ and $T_i$, there are two possible values
for $\rho_i$ and $P_i$ unless $E > \frac{3}{2} T_i$
which has only one possible value at $\rho_i > \rho_0$.
For a given temperature $T_i$,
which is lower than the critical temperature $T_c$,
there are two spinodal points; one at the liquid side with $\rho_l$ and $E_l$
and another one at the gas side with $\rho_g$ and $E_g$.
We now discuss, assuming equilibrium thermodynamics,
different possible behaviors of the remnant system
in various energy regions for a given initial temperature $T_i$.

If $E$ is low enough to be in the liquid side of the
spinodal stability region of the $T_i$ isothermal line, i.e., $E < E_l$,
then the system is either compressed or expanded (due to its nonzero pressure).
However, the system maintains uniformity since it is in the spinodal
stability region and the temperature should change to keep $E$ conserved.
For $E \le E_l$, the system will eventually equilibrate to the zero pressure
liquid phase at $T_f$ with energy $E$.
If $E = E_l$ with $\rho_l$, the system equilibrates at $T_f = T_l$.
Here $T_l$ is the temperature at which the energy
of the $P = 0$ point is $E_l$
and it is obvious that $T_i < T_l < T_{max}$ (see Fig.\ref{eosfig}).
In this case no break up has occured and
the system equilibrates at temperature $T_l$.

If now $E$ is such that $E_l < E < E_g$,
then the system is either in the spinodal instability
region of $T_i$ or is in a highly compressed dense liquid phase.
If the system stays uniform long enough, then
it will be compressed or expanded depending on the pressure
until it reaches $T_f$ with $P = 0$ and energy $E$.
For this situation, $T_f > T_l$.
However, for $\rho_i < \rho_0$, the system would not have a uniform
density due to the spinodal instability.
Instead, the system will break up, 
thereby becoming non-uniform and composed of two parts
with the total energy $E$ conserved.
One part will be a liquid phase with $E_l$ and $\rho_l$ and
another part is a gas phase with $E_g$ and $\rho_g$
at temperature $T_i$.
We may suppose that the system breaks up (fragments) at $\rho <\rho_l$
and the constituents clusterize at $\rho >\rho_g$ which becomes
the freeze out densities. 
The energy may flow from the dense part to the dilute part,
appearing as liberation energy and the thermal energy
of the fragments during this dynamical fragmentation process.
If we assume that the dynamical rearrangement process is faster
than the thermalization process in the spinodal instability region,
then the temperature stays at $T_i$ while it fragments.
However, the clusters of liquid phase will eventually equilibrate
internally to $E_l$, $T_l$ and zero pressure
without any further clusterization or fragmentation.
The gaseous system which is composed of nucleons and clusters
(each cluster behaving as a particle)
would then equilibrate thermally with the internal state of clusters at $T_l$,
and then it expands to an infinite volume due to the positive pressure.
The $T_l$ becomes the freeze out temperature of the gas.
Here $T_f = T_l$ and $E = \alpha E_l + (1 - \alpha) E_g$
but $\rho_i \ne \alpha \rho_l + (1 - \alpha) \rho_g$.

If the energy is even higher such that $E_g < E < E_n$,
then the system is either in the uniform
gas phase with density $\rho_i$ lower than $\rho_g$ or in highly compressed
very dense liquid phase with positive pressure.
If $\rho_i < \rho_g$, the system expands to an infinite volume
to achieve zero pressure without any clusterization.
If the system were initially in a very dense liquid phase $\rho_i > \rho_0$,
then it can reach a zero pressure point with temperature $T_f$ within
the liquid phase of the whole system (for $E < E_z$).
Another possibility is an expansion from the dense phase that
over shoots into the spinodal instability region or
to the gas phase region of the temperature $T_i$.
If it ends up in the gas phase region, i.e., $\rho < \rho_g$, then
it will further expand to infinite volume since $P > 0$.
If the original expansion reaches the spinodal instability region,
then the system would break up into a mixture of liquid phase and gas phase.
However since the energy $E$ is higher than $E_g$, the liquid phase
should share more energy than $E_l$ to thermally equilibrate
the internal temperature of each cluster of liquid phase and
the temperature of thermal motion of the gaseous fragments.
Thus the freeze out temperature $T_f$ now become higher than $T_l$
and $T_f$ increases as $E$ increases further for $E > E_g$.
Here $E > \alpha E_l + (1 - \alpha) E_g$
and the exceeded energy of this inequality makes $T_f > T_l$.

If the energy is higher than $E_n$,
then the system would not have any region of spinodal instability.
Thus the only possible phase is the gas phase (due to $P > 0$)
made of nucleons without any cluster of liquid phase.

These processes are represented by each broken line for each
temperature $T_i$ in Fig.\ref{phstrnf}.
If $\rho_i > \rho_l$ with $E < E_l$, then the whole system becomes
a uniform liquid at $T_f$ having zero pressure and energy $E$.
The low energy -- low temperature part of Fig.\ref{phstrnf} represents
the zero pressure point of a liquid phase which is very close
to $E^* = T^2 / \epsilon_0$, the low $T$ limit of Eq.(\ref{etpa}).
For the nuclear matter equation of state shown in Fig.\ref{eosfig},
$E^*_l = 2.1$ MeV (point $s$) and the $T_l \approx 5$ MeV for $T_i = 0$.
The liquid side spinodal density $\rho_l$
becomes the lower limit of the spinodal stability region
of the equilibrated Fermi liquid system
and $T_l \approx 5$ MeV becomes the break up temperature \cite{prcano}
at which a nuclear system fragments into two clusters in average.
The thermal properties of a nuclear system
below $E \approx 2$ MeV and $T \approx 5$ MeV
are hard to be extracted from data \cite{predata}.
If $\rho_i < \rho_g$ with $E > E_g$,
then the whole system becomes a uniform gas at the freeze out
temperature $T_f$ and density $\rho_g$.
The very high temperature -- high energy part (which is not shown in
Fig.\ref{phstrnf}) represents a gas system without any clusters
and the high $T$ limit of Eq.(\ref{etpa}) applies to this region.

If the density is $\rho_g < \rho_i < \rho_l$ then the system should be
a mixture of liquid of $\rho > \rho_l$ and gas of $\rho < \rho_g$
at temperature $T_f$ 
with zero pressure.
The spinodal densities $\rho_l$ and $\rho_g$ of temperature $T_i$ become
the break up (fragmentation) density of clusters and the freeze out
(clusterization) density of gas respectively at $T_f$.
For $E < E_g$, there is not enough energy for the system to become
a uniform gas, and for $E > E_l$, there is too much energy for the
system to maintain a uniform liquid phase as a whole.
The flat part of the caloric curve of Fig.\ref{phstrnf} represents
the spinodal instability region with $E_l < E < E_g$
between $\rho_l$ and $\rho_g$ where some parts form clusters of
liquid phase of $T_f = T_l$ with $E_l$.
In turn, this system of clusters forms a gas system at $T_l$,
with freeze out density $\rho_g$.
The higher temperature -- higher energy part of Fig.\ref{phstrnf} represents
the liquid-gas mixed phase of temperature $T_f > T_l$ with $E > E_g$.
Here the liquid phase has zero pressure with energy higher than $E_l$
and the gas phase has energy $E_g$ and
freeze out density of $\rho_g$.
For the nuclear matter results of Fig.\ref{eosfig},
$E^*_g = 16.15$ MeV for zero temperature.
This behavior of $E$--$T$ caloric curve can be represented
by the results Eq.(\ref{etpa}) with two different freeze out
densities, $\rho_l$ for liquid phase and $\rho_g$ for gas phase.
As can be seen from Fig.\ref{phstrnf}, the data of Ref.\cite{prldata}
can be fit with $T_i = 0 \sim 1.5$ MeV
and with $a_B = \epsilon_0 = 10$ MeV.

For a given energy $E$, the temperature $T_i$ of system may have any
value consistent with $0 \le T_i \le T_{max}$.
Various $T_i$ cause various $T_f$ and thus we need to superpose
these various temperatures with some weight.
For smaller beam energy the temperature range is smaller and the average
is not much different than taking a single temperature.
A superposition of $T_i$'s with $0 \le T_i \le 2$ MeV
and with $a_B = \epsilon_0 = 10$ MeV fits the data
of Ref.\cite{prldata} of 600 MeV/A Au + Au collisions
well (the thin solid line which
is similar to the thick solid line for $T = 1$ MeV
using the spinodal densities $0.677 \rho_0$ and $0.015 \rho_0$ as the
freeze out densities of liquid and gas phases respectively).
For a collision with larger energy, averaging over
a larger range of temperature gives a smooth curve of $E$--$T$.
The superposition $0 \le T_i \le 8$ MeV with $a_B = \epsilon_0 = 6$ MeV
fits the data of Ref.\cite{predata} of 1 GeV Au + C collisions
(the smooth thin solid line).
This range of $T_i$ is compatible with $T_i = 8.7$ MeV at critical
multiplicity given Ref.\cite{predata}.

The above analysis indicates that the liquid to gas phase transition 
(break up of nucleus) starts to occure at the temperature and
energy at which the pressure is zero and the energy is the same
as the liquid side spinodal point of zero temperature.
It also indicates that a low energy collision would show a clearer
first order phase transition than a high energy collision once the
excitation energy is high enough to reach the zero temperature
spinodal energy.
A high energy collision with $T_{max} > T_c$ is required in study
of critical behavior.
The different set of values of $a_B$ and $\epsilon_0$ used in the fits for
the low and the high beam energies show that the multifragmentation
process with lower beam energy is better applicable than the higher energy
reaction in extracting information for infinite nuclear matter.
The lower energy reaction gives larger size fragments and
thus closer to infinite system.
We need further study to check if the dynamical rearrangement process is
faster than the thermalization process in the spinodal instability region
and the other way around for spinodal stability region.

This work was supported in part by the Korea Science and Engineering
Foundation under Grant No. 951-0202-048-2, in part by the Kyung Hee
University under Grant No. 2U0195074, in part by the Ministry of
Education, Korea through Basic Science Research Institute Grant
No. 96-2422, and in part by the DOE Grant No. 4-25175.

%
%
%
%

%
%

\end{document}